\documentclass{jfm}

\usepackage{graphicx}
\usepackage{amsmath,amssymb,amstext} 
\usepackage{pifont}
\usepackage{xcolor}
\usepackage{framed}

\usepackage{epstopdf}
\usepackage[]{hyperref}
\usepackage{color,soul}

\usepackage{subcaption}
\usepackage{enumerate}

\usepackage[letterpaper]{geometry}

\ifCUPmtlplainloaded \else
 \checkfont{eurm10}
 \iffontfound
    \IfFileExists{upmath.sty}
     {\typeout{^^JFound AMS Euler Roman fonts on the system,
                  using the 'upmath' package.^^J}%
      \usepackage{upmath}}
     {\typeout{^^JFound AMS Euler Roman fonts on the system, but you
                  dont seem to have the}%
      \typeout{'upmath' package installed. JFM.cls can take advantage
                of these fonts,^^Jif you use 'upmath' package.^^J}%
     }
 \else
 \fi
\fi

\ifCUPmtlplainloaded \else
 \checkfont{msam10}
 \iffontfound
    \IfFileExists{AMSsymb.sty}
     {\typeout{^^JFound AMS Symbol fonts on the system, using the
                'amssymb' package.^^J}%
      \usepackage{amssymb}%
      \let\le=\leqslant  
      \let\ge=\geqslant  
     }{}
 \fi
\fi

\ifCUPmtlplainloaded \else
 \IfFileExists{amsbsy.sty}
    {\typeout{^^JFound the 'amsbsy' package on the system, using it.^^J}%
    \usepackage{amsbsy}}
    {\providecommand\boldsymbol[1]{\mbox{\boldmath $##1$}}}
\fi

\usepackage[authoryear]{natbib}
\setcitestyle{authoryear,round}

\newcommand{\pderiv}[2]{\frac{\partial #2}{\partial #1} }
\newcommand{\deriv}[2]{\frac{\mathrm{d}#2}{\mathrm{d} #1} }

\newcommand{\dx}{\mathrm{d}x}
\newcommand{\dz}{\ \hbox{d} z}

\newcommand{\dA}{\ \hbox{d} A}

\newcommand{\Vavg}[1]{<#1>_V }
\newcommand{\Havg}[1]{\overline {#1}}

\newcommand{\TB}{T_B}
\newcommand{\TBT}{T_{max}}
\newcommand{\TBTV}{0.37}

\newcommand{\TU}{T_U}

\newcommand{\TMD}{\tilde T_{md} } 
\newcommand{\Tf}{\tilde T(\tilde z=H)}
\newcommand{\mT}{\overline T}

\newcommand{\wT}{\overline{w^\prime T^\prime}}

\newcommand{\wRho}{\overline{w^\prime \rho^\prime}}

\newcommand{\delST}{\delta_{St}}
\newcommand{\delBL}{\delta_{BL}}

\newcommand{\TKE}{\hbox{TKE}}
\newcommand{\Ra}{\hbox{Ra}_0}
\newcommand{\RaL}{\hbox{Ra}_{Lin}}
\newcommand{\RaE}{\hbox{Ra}_{eff}}
\newcommand{\RaEM}{\hbox{Ra}_{max}}

\begin{document}
\title[The Cooling Box Problem]{The Cooling Box Problem: Convection with a quadratic equation of state }
\author{Jason Olsthoorn\thanks{Email address for correspondence: Jason.Olsthoorn@ubc.ca}, Edmund W. Tedford, and Gregory A. Lawrence}

\affiliation{Department of Civil Engineering, University of
British Columbia, 2002-6250 Applied Science Ln, Vancouver, BC V6T 1Z4}


\maketitle

\begin{abstract}

We investigate the convective cooling of a fluid with a quadratic equation of state by performing three-dimensional direct numerical simulations of a flow with a fixed top-boundary temperature, which is lower than the initial fluid temperature. We consider fluid temperatures near the density maximum, where the nonlinearity is expected to be important.
When the equation of state is nonlinear, the resultant vertical transport of heat is fundamentally different and significantly lower than the predictions derived for a linear equation of state. Further, three dimensionless groups parameterize the convective system: the Rayleigh number ($\Ra$), the Prandtl number (Pr), and the dimensionless bottom water temperature $(\TB)$. We further define an effective Rayleigh number ($\RaE = \Ra \ T^2$), which is equivalent to the traditional Rayleigh number used with a linear equation of state. We present a predictive model for the vertical heat flux, the top boundary-layer thickness, and the turbulent kinetic energy of the system. We show that this model agrees well with the direct numerical simulations. This model could be used to understand how quickly freshwater lakes cool in high latitude environments. 
\end{abstract}


\section{Introduction} \label{Sec::Introduction}



An important feature of freshwater lakes is that they have a nonlinear equation of state (EOS). This nonlinear EOS is nearly quadratic with temperature near the temperature of maximum density, which is above the freezing temperature of the water ( e.g. $\TMD \approx 3.98^\circ $C for distilled water at atmospheric pressure). The significance of this nonlinearity for lakes has been recognized for over a century \citep{Whipple1898}. As a result of the density maximum, cooling the surface of a water body that has a mean internal temperature below $\TMD$ will stabilize the water column resulting in the characteristic reverse temperature stratification found during the winter months in temperate lakes \citep{Farmer1975}. Stratification with temperatures on opposite sides of $\TMD$ will lead to cabbeling (nonlinear mixing), which has important implications for convection \citep{Farmer1981,Carmack1979,Couston2017,Couston2018}. Convection of this type is also relevant in other fields such as Arctic melt ponds \citep{Kim2018} and geologically sequestered carbon dioxide \citep{hewitt2013}. This paper aims to understand how thermal convection is altered near $\TMD$. 


The theoretical studies of such a system date back to \citet{Veronis1962}. He considered the linear stability of a fluid layer with fixed temperatures at the top and bottom boundaries; temperatures fixed on either side of $\TMD$. Shortly thereafter, \citet{Townsend1964} performed a complementary experimental study, again with fixed temperatures at the top and bottom. Both studies showed that the nonlinear EOS results in a stable upper layer above a convectively unstable lower layer, which will preferentially mix the lower-layer temperature stratification (similar to Figure \ref{fig::TankDiagram}(a) except with a bottom thermal boundary layer). Several studies have shown that convection can generate internal waves in the stable upper-layer, which are particularly relevant for astrophysical applications \citep{LeBars2015,Lecoanet2015}.
Recently, \citet{Toppaladoddi2017} and  \citet{wang2019} built on this previous work and performed two-dimensional and three-dimensional simulations of convection with a nonlinear EOS. Both studies highlighted that in addition to the traditional dimensionless parameters (the Rayleigh number and the Prandtl number) considered for this flow setup, the nonlinear EOS introduces an additional independent parameter. The additional nondimensional parameter quantifies the ratio of the temperature variation across the stable and unstable layers. We will show that a similar ratio is important here. In each of these studies, the top and bottom temperatures are fixed, and the results focus on the statistical steady state.

Most lakes do not reach a steady-state, but warm and cool throughout the year. In addition, the dominant heat loss in these freshwater systems is through the water surface with a relatively insulated bottom. Motivated by these considerations, we study a box of warm fluid ($\tilde T>\TMD$) cooled from above ($\tilde T<\TMD$). Here, unlike the previous studies, the lower boundary is insulating, and the temperature stratification is transient. As is typical of these theoretical studies \citep{Veronis1962,Townsend1964,Olsthoorn2019}, we will consider an EOS that is quadratic with temperature, which approximates the full EOS of \citet{ChenMillero1986} for temperatures below 10 $^\circ$C. In this paper, we will refer to a cooling of the surface, though the problem is symmetric for a box of cold water that is heated from above, at least for a quadratic EOS assumed here. We want to understand how convection and the resultant heat transport is changed in the presence of this nonlinear density relationship, near $\TMD$. In particular, we address the three following questions:
\begin{enumerate}
        \vspace{6pt}
    \item Does the nonlinear equation of state affect the vertical transport of heat within the water?
        \vspace{2pt}
    \item What parameters determine the heat flux out of the water surface?
        \vspace{2pt}
    \item Can we predict the vertical heat transport and kinetic energy produced by the turbulent convection?
    \vspace{6pt}
\end{enumerate}   

We will begin, in \S\ref{sec::ExpSetup}, with a discussion of the numerical methods used in this paper. Section \S\ref{sec::NumSim} is an analysis of the three-dimensional direct numerical simulations, highlighting that the vertical transport of heat is significantly different for a nonlinear EOS than that predicted for a linear EOS. Section \ref{sec::ScalingLaws} discusses the relevant parameters that control the heat flux and present a predictive model for this system. We show that the model agrees well with the data from the numerical simulations in \S\ref{sec::ModelCollapse}.  Finally, we conclude in \S\ref{sec::Conclusions}.

\section{Numerical Setup} \label{sec::ExpSetup}

\begin{figure}
\centering
\includegraphics[width=\textwidth]{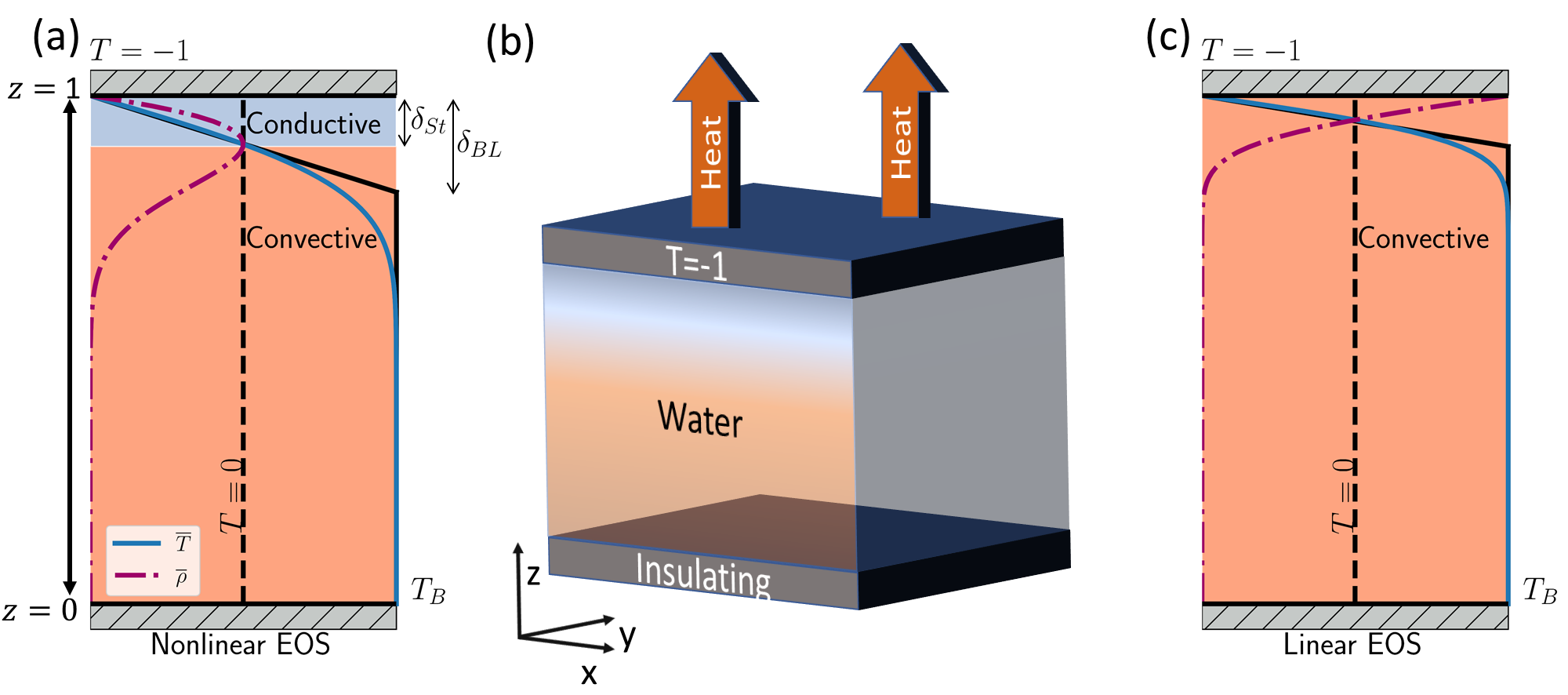}
\caption{(a) Representative mean temperature (blue line) and density (magenta dashed-dot line) profiles with a nonlinear equation of state. Note the presence of an upper stable thermal boundary-layer (blue). The model's piecewise-linear profiles (equation \eqref{eqn::meanT}) are also included as a solid black line. (b) A diagram of the numerical domain. (c) Comparative mean temperature and density profiles for a linear equation of state. }
\label{fig::TankDiagram}
\end{figure}

We consider a body of water that is insulated from below and cooled from above. We restrict our analysis to freshwater
, where the temperature of maximum density $\TMD$ is above its freezing temperature. Figure \ref{fig::TankDiagram}(b) is a schematic of the numeric domain of interest. We ignore the effects of salinity, pressure, and higher-order terms in the equation of state (EOS) such that the density ($\tilde \rho$) is given
\begin{align}
\tilde \rho &= \tilde \rho_0  -  C_T\left(\tilde T -  \TMD\right)^2.
\end{align}
Here, $C_T$ is a constant and $\tilde \rho_0$ is the density at $\tilde T = \TMD$. 

We are interested in parameterizing the convective heat flux through the top boundary. As such, a natural time scale for this setup is the diffusive timescale of heat $\tau_{\kappa}$ over the domain height $H$. That is,
\begin{align}
 \tau_{\kappa} = \frac{H^2}{\kappa},
\end{align}
where $\kappa$ is the diffusivity of heat.
We then nondimensionalize the spatial coordinates ($\tilde{ \mathbf{x}}$), the fluid velocity ($\tilde{\mathbf{ u }}$), time ($\tilde t$) and temperature ($T$) as, 
\begin{gather}
\mathbf{x} = \frac{\tilde{ \mathbf{x}}}{H} , \qquad \mathbf{u} = \frac{\tilde{\mathbf{ u  }} H}{\kappa}, \qquad  t = \frac{\tilde{ t}}{\tau_{\kappa}}, \qquad T = \frac{\tilde T - \TMD}{\Delta \tilde T}. 
\end{gather}
Boldface variables denote vector quantities and $\Delta \tilde T = \TMD - \Tf$. As an aside, we would like to highlight that, throughout the paper, we have dropped factors of depth Lz $=1$. While we believe that this is a reasonable simplification, we highlight this convention to avoid any confusion for the reader. 

The equations of motion for this flow are the Navier-Stokes equations under the Boussinesq approximation. These equations are written
\begin{gather}
	\left( \pderiv{t}{} + \mathbf{u} \cdot \nabla \right)\mathbf{u} = - \nabla P  +  \Ra \hbox{ Pr } T^2 \mathbf{\hat k} + \hbox{Pr} \nabla^2 \mathbf{u},
	\label{eqn::momentum}\\
	\left( \pderiv{t}{} + \mathbf{u} \cdot \nabla \right) T = \nabla^2 T, \label{eqn:Temp}	\\
	\nabla \cdot \mathbf{u} = 0. \label{eqn::DivFree}
\end{gather}
We have defined the Rayleigh number ($\Ra$) and Prandtl number (Pr) as 
\begin{gather}
\Ra = \frac{ g C_T \Delta \tilde T^2}{\rho_0}\frac{H^3}{\kappa \nu}, \qquad \hbox{Pr} = \frac{\nu}{\kappa}, 
\end{gather}
where $\nu$ is the molecular viscosity of water, and $g$ is the gravitational acceleration.  The value of the Prandtl number for freshwater varies from $\Pr\approx 7$ at $\tilde T = 20 ^\circ $C to $\Pr \approx 13.4$ at $\tilde T = 0 ^\circ$C, largely due to variations in $\nu$. Incorporating the functional dependence of $\nu$ on $T$ is outside of the scope of this paper. \cite{Hay2019} performed convective simulations with an evolving Pr, in a different context, that do not show significant changes to the vertical heat flux from the constant Pr case. Future work will discuss the role of the changing viscosity at low temperatures. In this paper, we will assume a constant $\nu$, and prescribe $\Pr=9$. 

We enforce a fixed temperature on the surface and an insulating bottom condition. That is  
\begin{gather}
T\bigg|_{z = 1} = -1, \qquad \frac{\partial T}{\partial z}\bigg|_{z=0} = 0,
\end{gather}
along with no-slip top and bottom velocity boundary conditions.


%
%
%
%
As we will see below, the well-mixed bottom-water temperature $T_B$ is an important parameter in this system. As illustrated in Figure \ref{fig::TankDiagram}(a), while the system is convectively unstable, the temperature profile below the top boundary layer is nearly uniform.
In the numerical simulations, we will determine $\TB$ by fitting the horizontally averaged temperature profile to an $\hbox{erf}$ function. That is, \begin{align}
    \Havg{T} \approx \left( 1 + \TB\right) \hbox{erf} \left( \frac{z-1}{\eta}\right) - 1,
\end{align}
where $\eta\approx \delBL$ (defined below) is a fit parameter and $\Havg{\hbox{ }\cdot \hbox{ }} = \frac 1 {\hbox{Lx Ly}} \int \cdot \ \dx\hbox{ d}y$. Over time, $\TB$ will decrease, whereas $\Ra$, by definition, will remain fixed.

In most cases, we selected the initial fluid temperature such that there was
 an initial symmetry between the bottom water temperature ($\TB=1$) and the top boundary condition about $T=0$. For reference, for a water depth of $H=0.05 \text {m}$ with $\Tf =0^\circ \text{C}$, the initial water temperature would be $\tilde T \approx 8 ^\circ $C and $\Ra\approx 8\times10^{5}$.
Thus, the simulations presented in this paper (see table \ref{Table::NumParams}) are on the scale of potential laboratory experiments. In the lowest $\Ra$ case, we selected $\TB (t=0) = 2$ so that the initial instability grew significantly quicker than the background diffusion.

While the top boundary temperature will remain fixed ($T(x,y,z=1,t)=-1$), $\TB$ will cool throughout the numerical simulations. While $\TB>0$, there exist two sub-domains to the mean temperature stratification: an upper stable stratification of depth $\delST$ where $\Havg{T}<0$ and a lower hydrostatically-unstable stratification where $\Havg{T}>0$. 
Figure \ref{fig::TankDiagram}(a) is a plot of a representative temperature profile within the fluid domain. 
The total top boundary-layer thickness between the well-mixed uniform temperature and the upper boundary is $\delBL>\delST$ (see Figure \ref{fig::TankDiagram}(a)).

Before continuing, we highlight that for a linear EOS, $\TMD$ cannot be internal to the fluid domain, and an upper stable layer cannot form. As such, for a linear EOS, we would expect a temperature profile to resemble that in Figure \ref{fig::TankDiagram}(c).

\subsection{Numerical Implementation}
We performed direct numerical simulations using SPINS \citep{Subich}. SPINS solves the Navier-Stokes equations under the Boussinesq approximation using pseudospectral spatial derivatives and a third-order time-stepping scheme. As the top boundary-layer controls the dynamics of the initial instability and the subsequent convection, we implement a Chebyshev grid in the vertical, which clusters grid points at the domain boundaries. We assume periodic horizontal boundary conditions, implemented with Fast Fourier Transforms. 

We performed five numerical simulations with a nonlinear EOS at different Rayleigh numbers. We performed one additional simulation with a linear EOS for comparison. The details of these numerical simulations are provided in Table \ref{Table::NumParams}. In all six cases, the Rayleigh number was large enough for the system to become unstable (See Appendix \ref{App::LinStab} for the linear stability analysis). We initially perturb the three velocity components with a random perturbation sampled from a Normal distribution scaled by $10^{-2}$. The numerical resolution (Nx $\times$ Ny $\times$ Nz) was selected, such that $\max  \frac{\Delta x}{\eta_B}<3$, where we compute the Batchelor scale 
$\eta_B = \left(\Havg{\varepsilon}\right)^{-\frac{1}4} \Pr^{-\frac{1}2}$, 
for viscous dissipation rate $\varepsilon$ (see equation \eqref{eqn::TKEBudget}) and horizontal grid spacing $\Delta x$. The vertical grid is clustered towards the boundaries and $\max  \frac{\Delta z}{\eta_B}<\max  \frac{\Delta x}{\eta_B}$, in all cases. This resolution criterion is similar to that employed in \citet{Hay2019,Kaminski2019,Olsthoorn2019,Smyth2000}. The resolution sufficiency is highlighted in Appendix \ref{App::Res}.

\begin{table*}
\centering
\begin{tabular}{c||c|cc|ccc|c}
Case & EOS &  Domain Size  & Resolution & $\Ra$ & Pr & $T_B$  & $\max \frac{\Delta x}{\eta_B}$\\
 &&  (Lx $\times$ Ly $\times$ Lz) & (Nx $\times$ Ny $\times$ Nz) &&& $(t=0)$ &\\
\hline
1 &\  Quadratic\  & $ 4\times 4 \times 1$ & $256 \times 256 \times 128$ & $9.0\times 10^4$ & 9 & 2 & 1.5 \\ 
2 &\  Quadratic\  & $ 4\times 4 \times 1$ & $256 \times 256 \times 256$ & $4.5\times 10^5$ & 9 & 1 &  1.8\\ 
3 &\  Quadratic\  & $ 4\times 4 \times 1$ & $256 \times 256 \times 256$ & $9.0\times 10^5$ & 9 & 1 & 2.1 \\ 
4 &\  Quadratic\  & $ 4\times 4 \times 1$ & $512 \times 512 \times 256$ & $4.5\times 10^6$ & 9 & 1 & 1.6 \\ 
5 &\  Quadratic\  & $ 4\times 4 \times 1$ & $512 \times 512 \times 256$ & $9.0\times 10^6$ & 9 & 1 & 2.2 \\ 
\hline
6 & Linear& $ 4\times 4 \times 1$ & $256 \times 256 \times 256$ & $9.0\times 10^5$ & 9 & 1 & 1.9 \\ 

\end{tabular}
\caption{Table of the parameters for each numerical simulation.}
\label{Table::NumParams}
\end{table*}


\section{Simulation Results}\label{sec::NumSim}

In this section, we describe both the qualitative and quantitative dynamics of the numerical simulations as they relate to the transport of heat within the fluid domain. In particular, we show that the convection is self-similar in $\TB$ for the range of Rayleigh numbers considered. Our discussion is focused on a single representative case: Case 3, Ra$=9.0\times10^5$. The results are similar for the other simulations, except where otherwise noted.

\begin{figure}
\centering
\includegraphics[width=\textwidth]{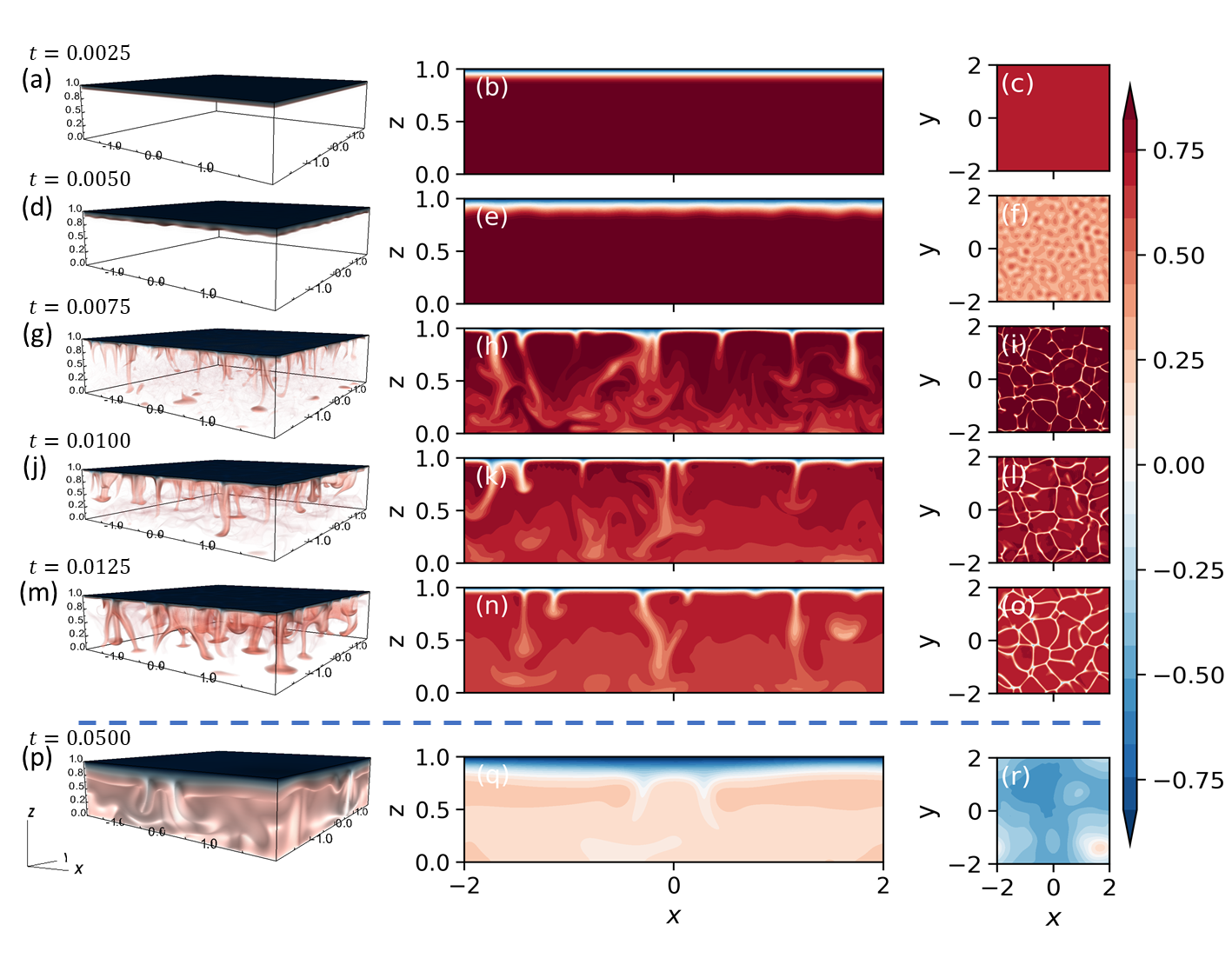}
\caption{Snapshots of the temperature field for Case 3: $\Ra=9.0\times 10^5,\ \Pr=9$. The volume plots \{(a),(d),(g),(j),(m),(p)\} encapsulate the three-dimensional structure of the flow field. The middle figure panels \{(b),(e),(h),(k),(n),(q)\} are contour plots of vertical ($x-z$) temperature slices at $y=-1$. Similarly, The right figure panels \{(c),(f),(i),(l),(o),(r)\} are contour plots of horizontal ($x-y$) temperature slices at $z=0.9$. Snapshots are given at $t=2.50\times 10^{-3}$(a)-(c), $t=5.00\times 10^{-3}$(d)-(f), $t=7.50\times 10^{-3}$(g)-(i), $t=1.00\times 10^{-2}$(j)-(l), $t=1.25\times 10^{-2}$(m)-(o), and $t=5.00\times 10^{-2}$(p)-(r). Note the jump in output times highlighted by the horizontal dashed line.}
\label{fig::RaPr1e5_Snapshots}
\end{figure}

Figure \ref{fig::RaPr1e5_Snapshots} contains snapshots of the temperature field for Case 3. The left column of Figure \ref{fig::RaPr1e5_Snapshots} contains plots of the temperature field at different times $t=\{2.50\times 10^{-3},\ 5.00\times 10^{-3},\ 7.50\times 10^{-3},\ 1.00\times 10^{-2},\ 1.25\times 10^{-2},\ 5.00\times 10^{-2}\}$. These plots were made with VisIt's \citep{VisIt} volume plot option that varies the transparency of the temperature field according to its value. The downwelling plumes are visible in the $X-Z$ slices (Figure \ref{fig::RaPr1e5_Snapshots} (middle)). Initially, the temperature stratification is linearly stable, and the temperature profile simply diffuses.  At $t=0.0025$ (Figure \ref{fig::RaPr1e5_Snapshots}(a)-(c)), the temperature stratification matches that predicted by pure diffusion. Once the top thermal boundary-layer has grown sufficiently, the system is unstable, and small perturbations to the velocity and temperature field will grow. These near-modal perturbations are visible at $t=5.00\times 10^{-3}$ (Figure \ref{fig::RaPr1e5_Snapshots}(d)-(f)). Once the perturbations grow large enough, they begin to merge, forming dense plumes that transport cold fluid to the bottom of the domain. The columnar plumes are highly variable in both space and time, and the resultant convection will mix the bottom fluid. Eventually (Figure \ref{fig::RaPr1e5_Snapshots}(p)-(r)), the bottom water temperature reduces sufficiently such that the vertical heat flux rapidly decreases.

Slices of the spanwise ($X-Y$) structure of the temperature field at z = 0.9 are presented in Figure \ref{fig::RaPr1e5_Snapshots} (right). While a complete analysis of this structure is outside of the scope of this paper, we highlight it here as it demonstrates that the spacing between the downwelling plumes is increasing over time. The horizontal length scale of the three-dimensional flow structure increases as $T_B\to0$, as viscous dissipation preferentially diffuses small scale motions. Eventually, the spacing between the plumes reaches the size of the domain, which limits the run time of these simulations. We have performed several simulations at different domain sizes and have determined that the present results are not domain size-dependent.

\begin{figure}
\centering
\includegraphics[width=\textwidth]{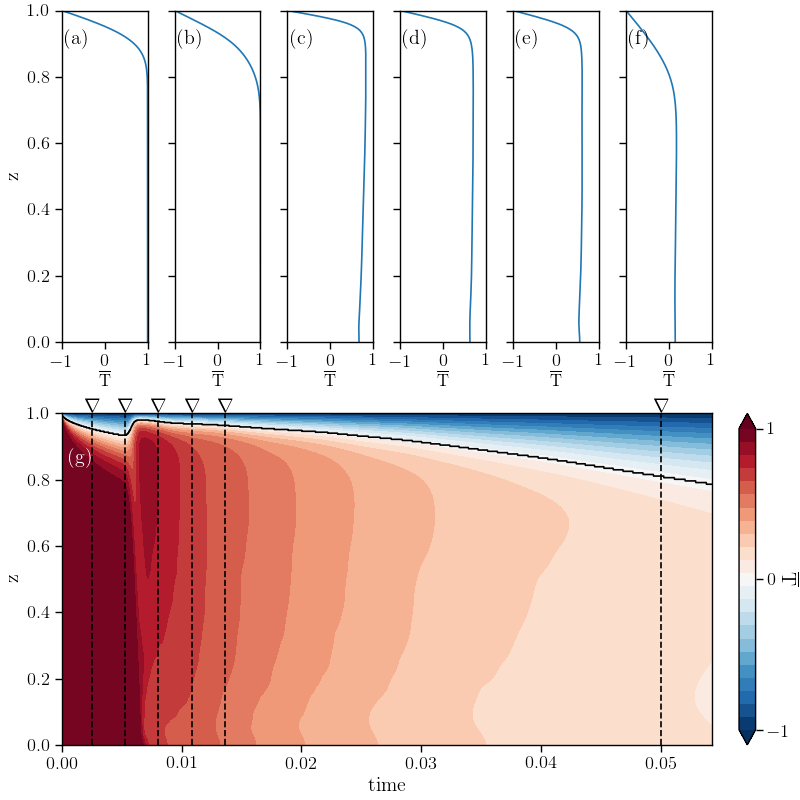}
\caption{The horizontally averaged temperature profiles are plot at the output time of the snapshots of Figure \ref{fig::RaPr1e5_Snapshots} ($t=2.50\times 10^{-3}$(a), $t=5.00\times 10^{-3}$(b), $t=7.50\times 10^{-3}$(c), $t=1.00\times 10^{-2}$(d), $t=1.25\times 10^{-2}$(e), and $t=5.00\times 10^{-2}$(f)). Contours of the horizontally averaged temperature $\left(\Havg{T}\right)$ over time are also plotted (g).
The vertical dashed lines in panel (g) represent the time of the vertical temperature profiles.
Data is shown for Case 3: $\Ra=9.0\times10^5$. }
\label{fig::RaPr1e5_HavgTemp}
\end{figure}

We highlight the mean structure of the temperature stratification in figure \ref{fig::RaPr1e5_HavgTemp}(a)-(f), which are plots of individual horizontally-averaged temperature profiles at times $t=\{2.50\times 10^{-3},\ 5.00\times 10^{-3},\ 7.50\times 10^{-3},\ 1.00\times 10^{-2},\ 1.25\times 10^{-2},\ 5.00\times 10^{-2}\}$ (identical to Figure \ref{fig::RaPr1e5_Snapshots}). 
The evolution of the temperature stratification is further highlighted in the contours of horizontally-averaged temperature over time (Figure \ref{fig::RaPr1e5_HavgTemp}(g)). These contours show that the boundary-layer depth increases over time and that the bottom water temperature decreases. Note that there is a weak thermal gradient near the bottom of the domain due to the solid boundary. This thermal gradient becomes weaker with increasing Ra, due to the increased energy within the system. 



\subsection{Vertical Heat Transport}\label{sec::HeatTransport}

Now that we have presented the essential features of the temperature evolution, we proceed to quantify the heat loss resulting from the thermal convection. The rate of heat loss of the water in the domain is  
\begin{gather}
    \deriv{t}{} \int_0^1 \int_A  T \dA \dz = -F A,
\end{gather}
where $F$ is the average outward heat flux at the domain surface, and $A=\hbox{Lx Ly}$ is the area of the water surface. The vertical transport of heat is approximately constant over the upper conductive boundary-layer.

The vertical heat flux within the convective layer is significantly greater than the diffusive flux. We define $\sigma$ as the ratio of average heat flux to the diffusive heat flux over the convective domain. That is, we write 
\begin{align}
    \sigma \approx \frac{F/\left( 1 - \delST\right)}{\left(T_U - T_0\right)/\left( 1 - \delST\right)} 
    = \frac{F}{T_U - T_0}, \label{eqn::Nu}
\end{align}
where $T_0$ is the temperature of maximum density within the domain. As we will see below, for a quadratic EOS, $\sigma$ is constant for a significant portion of the simulation time, which indicates the temperature decays exponentially. For a linear EOS, $T_0$ will be equal to the top-boundary temperature ($T_0=-1$). In this case, $\sigma$ is identical to the Nusselt number (Nu), which is the ratio of the average vertical heat flux to the diffusive heat flux across the entire domain. For the nonlinear EOS discussed here, where the temperature of maximum density is interior to the fluid domain, then $T_0 = 0$. 

 As an aside, it is worth noting that $F$ itself is a function of the temperature difference $\TU - T_0$. As the water becomes uniform, $\left( \TU - T_0\right) \to 0$, the vertical flux $F\to 0$ such that $\sigma$ is bounded.

As a simple model, we first consider the scenario where the water column is uniformly mixed to some temperature $\TU$. If $\sigma$ were a constant $\sigma_0$, then 
\begin{gather}
    \deriv{t}{} \TU A  = -\sigma_0 \cdot \left(\TU - T_0\right) A \implies \TU = B \exp \left( -\sigma_0 t\right) + T_0,
\end{gather}
where $B$ is a constant of integration. That is, $\TU$ decays exponentially in time. In principle, there is no \emph{a priori} reason to suspect that $\sigma$ will be constant and, in general, it is not. However, as we will show below, for the initial convective evolution, $\sigma$ is constant, and the temperature will decay exponentially. 



Before continuing, we return to the first of the main questions we are trying to answer in this paper. Does the nonlinear equation of state affect the vertical transport of heat out of the domain? We performed a single numerical simulation with a linear EOS with $\Ra = 9.0 \times 10^5$ and $\Pr=9$. The boundary and initial conditions remain unchanged. For a linear EOS, there are only two free parameters: a Prandtl number (Pr) and a time-dependent Rayleigh number. The relevant Rayleigh number for a linear EOS includes the density difference across the domain $\Delta \rho$ as
\begin{gather}
    \RaL = \frac{g \Delta \rho}{\rho_0} \frac{H^3}{\kappa \nu} = \Ra \left(1 + \TB\right).
\end{gather}
Here, we have related the constant Rayleigh number ($\Ra$) defined in this paper with a more traditionally defined time-varying Rayleigh number ($\RaL$), used with a linear EOS. The equivalent effective Reyleigh number ($\RaE$) for the quadratic EOS is given:
\begin{align}
    \RaE = \Ra \TB^2. 
\end{align}
We will return to this in our discussion of $\sigma$ below.

For a linear equation of state, it is empirically determined that  
\begin{gather}
    \hbox{Nu}_{Lin} \propto \RaL^n. \label{eqn::LinScaling}
\end{gather}
While significant controversy persists over the exact value of $n$ \citep[for example, see][]{Plumley2019}), we will specify $n=0.28$ as it best fits the data discussed below.

\begin{figure}
\centering
\includegraphics[width=14cm]{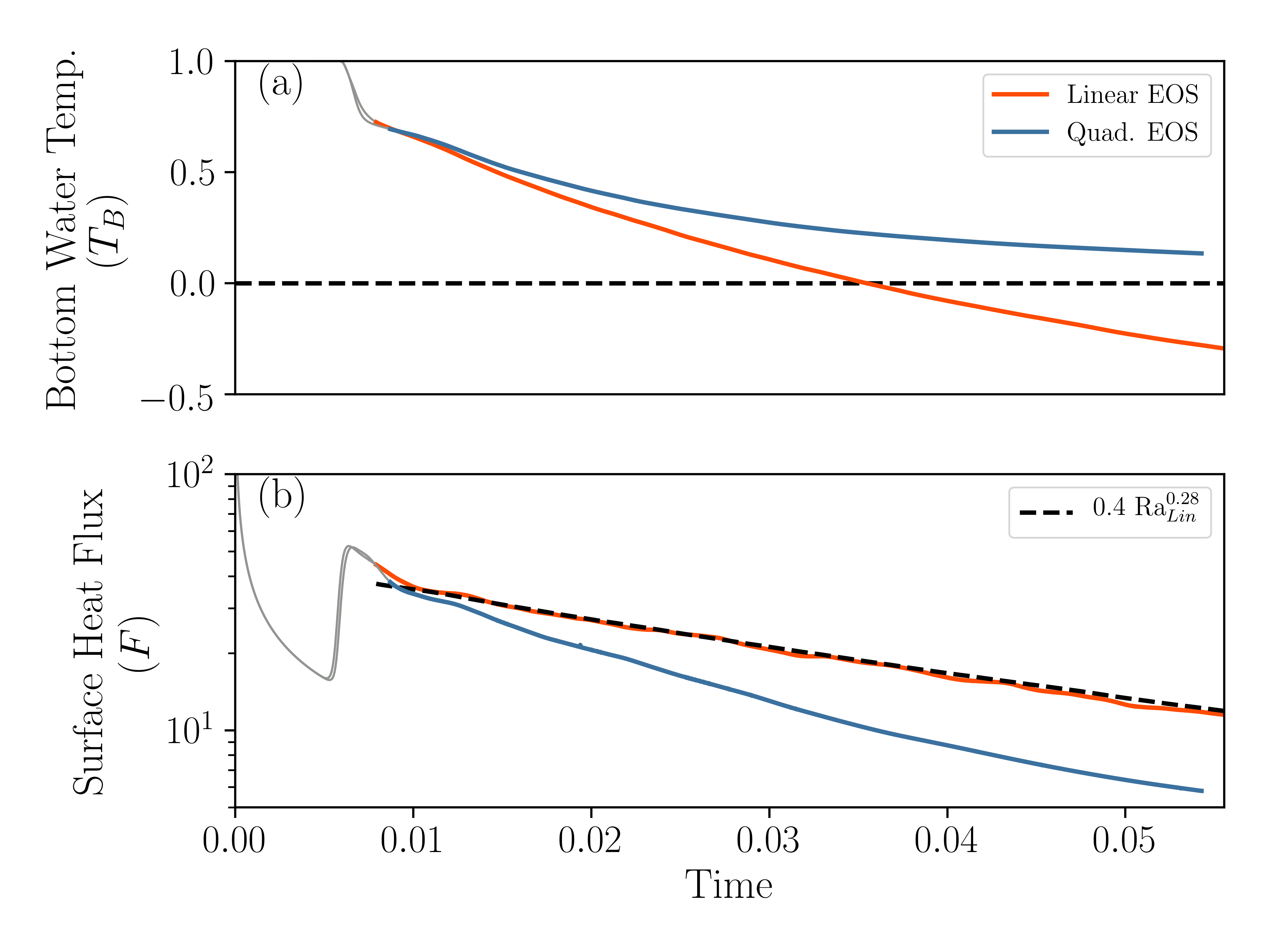}
\caption{A comparison plot of (a) $\TB$ and (b) $F$ versus time between convection with a linear or a quadratic equation of state. The temperature of maximum density is included as a horizontal dashed line in (a). The scaling \eqref{eqn::LinScaling} is included as a dashed line in panel (b). In both cases, $\Ra=9.0\times 10^5$. We plot in grey the initial (diffusive) transition period.}
\label{fig::FluxComparison} 
\end{figure}

 Figure \ref{fig::FluxComparison} is a comparison plot of the (a) temperature and (b) vertical heat flux as a function of time, a comparison between a linear or a quadratic EOS. The scaling \eqref{eqn::LinScaling} is included in panel (b). Here, and for the rest of the paper, we plot in grey the initial transition period before reaching a quasi-equilibrium  (where the vertical buoyancy flux approximately balances viscous dissipation).  We observe that the curvature in the EOS fundamentally changes the temperature evolution of the system over time. The surface heat flux is significantly greater for the linear EOS, resulting in the bottom water temperature $\TB$ decaying much faster for a linear EOS. For the quadratic EOS, the presence of a temperature of maximum density also restricts the convective mixing such that $\TB\to 0$ (note that molecular diffusion will eventually reduce $\TB\to -1$ as $t\to \infty$). If we return to the dimensional example of a 0.05 m deep container with a surface temperature at $0\ ^\circ C$, then by $t=0.05$ ($\approx 15$ minutes) there is a $1.5 \ ^\circ $C degree difference between the internal temperatures of the linear and quadratic EOS; nearly a 40\% increase in the heat loss!

The Nusselt number dependency in equation \eqref{eqn::LinScaling} is inadequate to describe the temperature evolution for a quadratic EOS. In the next section, we derive a model for the vertical heat flux ($F$) and turbulent kinetic energy density ($\TKE$) of convection with a nonlinear EOS. We will show that the convection is fundamentally dependent on three independent parameters, as opposed to the two needed with a linear EOS.

\section{Scaling Laws} \label{sec::ScalingLaws}

We begin with a Reynolds decomposition of the temperature and velocity field into a mean temperature profile and fluctuations from it, where
\begin{gather}
\mathbf{u} = \mathbf{0} + \mathbf{u}', \qquad T = \mT + T'. \label{eqn::ReyDecomp}
\end{gather}
In the numerical simulations, we will take $\left(\Havg{\hbox{ }\cdot\hbox{ }}\right)$, the horizontal average through the domain. 
For this scaling analysis, we will simplify $\Havg{T}$ as a piecewise linear profile 
\begin{gather}
\mT = \begin{cases}
-1 - \frac{1 + T_B}{\delta_{BL}} \left(z - 1 \right) & z> 1 - \delBL,\\
T_B & z\le 1- \delBL,\\
\end{cases} \label{eqn::meanT}
\end{gather}
as in Figure \ref{fig::TankDiagram}(a). Note that $\delBL$ is the total transition thickness with $\delBL>\delST$.

\subsection{Boundary-Layer Thickness $\delST$} 

The diffusion of heat through the top boundary is  
\[ F = - \pderiv{z}{\mT}\bigg|_{z=1} =  \frac{1 + \TB}{\delBL} = \frac{1}{\delST}.\]
That is, in this nondimensionalization, the boundary layer thickness solely determines the outward heat flux. As a reminder, $\delST$ is the top stable boundary-layer thickness between the top boundary and where $\Havg{T}=0$ (the temperature of maximum density).

Substituting the Reynolds decomposition \eqref{eqn::ReyDecomp} into the temperature evolution equation \eqref{eqn:Temp}, we derive the evolution equation for the mean temperature profile,
\begin{gather}
\pderiv{t}{\mT} = -\pderiv{z}{} \left( -\pderiv{z}{\mT} + \Havg{T'w'} \right). \label{eqn::Tbar}
\end{gather}
Balancing the heat fluxes within the domain and top boundary layer, we determine  
\begin{gather}
-\underbrace{\pderiv{t}{T_B} \left(1-\left(1 + T_B\right)\delta_{St}\right)}_{\hbox{Interior Cooling}} + \underbrace{\frac{\left(1 + T_B\right)^2}{2} \pderiv{t}{\delST}}_{\hbox{Increased $\delST$} } = \underbrace{\frac{1}{\delta_{St}}. }_{\hbox{Outward Heat Flux}} \label{eqn::St}
\end{gather}
Appendix \ref{App::DerivDelta} provides a derivation of \eqref{eqn::St}. We can see from equation \eqref{eqn::St} that a fraction of the heat loss is derived from the cooling of the interior fluid. The remainder of the heat loss is given by an increase in the boundary layer thickness $\delST$.  The surface heat loss is controlled by the boundary layer thickness $\delta_{St}$.

Building on \S \ref{sec::HeatTransport}, the average temperature within the domain is $\TB$ such that,
\begin{gather}
 \frac{d \TB}{dt} = -\sigma \TB.
 \end{gather}
As convection greatly enhances the vertical flux of temperature, we find that $\sigma\gg 1$. 
We will continue a discussion of $\sigma$ below. To leading order in $O\left( \frac 1 \sigma \right)$, the dominant balance between the interior cooling and the outward heat flux determines that
\begin{gather}
\delta_{St} \sim \frac{1}{\sigma T_B}, \qquad  F = \frac{1}{\delta_{St}} \sim \sigma T_B,  \qquad  \sigma\gg 1. \label{eqn::DelScaling}
\end{gather}


\subsection{Turbulent Kinetic Energy}

Similar to equation \eqref{eqn::Tbar}, the volume integrated turbulent kinetic energy density ($\TKE = \frac 12 \mathbf{u}^\prime \cdot \mathbf{u}^\prime $) evolution equation is written
\begin{gather}
\deriv{t}{\Vavg{\TKE}} = - \Ra \Pr \Vavg{w^\prime \rho^\prime } - \varepsilon, \qquad \varepsilon =\Pr \Vavg{ \nabla \mathbf{u}^\prime : \nabla \mathbf{u}^\prime }\label{eqn::TKEBudget}.
\end{gather}
The $(:)$ operator is the double dot product. Unlike the case of a linear EOS, the vertical buoyancy flux ($\Havg{w^\prime \rho^\prime}$) is not directly proportional to the vertical temperature flux ($\wT$). For a quadratic EOS, the vertical buoyancy flux is the sum of two components, 
\begin{gather}
\Havg{w^\prime \rho^\prime} = - 2 \Havg{w^\prime T^\prime} \ \Havg{T} - \Havg{w^\prime T^\prime T^\prime}. \label{eqn::BuoyFlux}
\end{gather}
Along with the mixing coefficient ($\Gamma$, the ratio of the vertical buoyancy flux to viscous dissipation), we define a buoyancy flux ratio ($\lambda$) to quantify the relative contribution of each buoyancy flux term. These are defined as  
\begin{gather}
\Gamma = \frac{ -\Ra \Pr \Vavg{\wRho}}{\varepsilon}, \qquad \lambda = \frac{-\Vavg{\Havg{w^\prime T^\prime T^\prime}}}{\Vavg{2 \Havg{w^\prime T^\prime} \ \Havg{T} }}.
\label{eqn::Lambda}
\end{gather}
%
%
Assuming self-similarity of TKE, we can show that  (see Appendix \ref{App::DerivTKE} for details)
\begin{gather}
\Vavg{\TKE} \sim \frac 12 \left( \Gamma^{-1} - 1 \right) \left( 1 - \Lambda \right) \Ra \Pr \TB^2, \qquad \sigma \gg 1, \label{eqn::TKEScaling}
\end{gather}
where $\Gamma$ is assumed constant and
\begin{gather}
    \Lambda = \frac{2}{\TB^2} \int_{\TB} T' \lambda(T') \ dT'.
\end{gather}
This model for the kinetic energy is consistent with the empirical Reynolds number scaling found in \cite{wang2019}. 

What parameters control the heat flux out of the water surface? The scaling laws \eqref{eqn::DelScaling} and \eqref{eqn::TKEScaling} depend on three undetermined coefficients: the decay rate $\sigma$, the buoyancy flux ratio $\lambda$, and the mixing coefficient $\Gamma$. These three coefficients are, as we will show, functions of the three parameters $\Ra, \Pr$ and $\TB$. In the next section, we will derive the functional form of $\sigma$, $\lambda$, and $\Gamma$, and show that the models provided here agree well with the simulated quantities.

\section{Model Comparison} \label{sec::ModelCollapse}


\begin{figure}
\centering
\includegraphics[width=10cm]{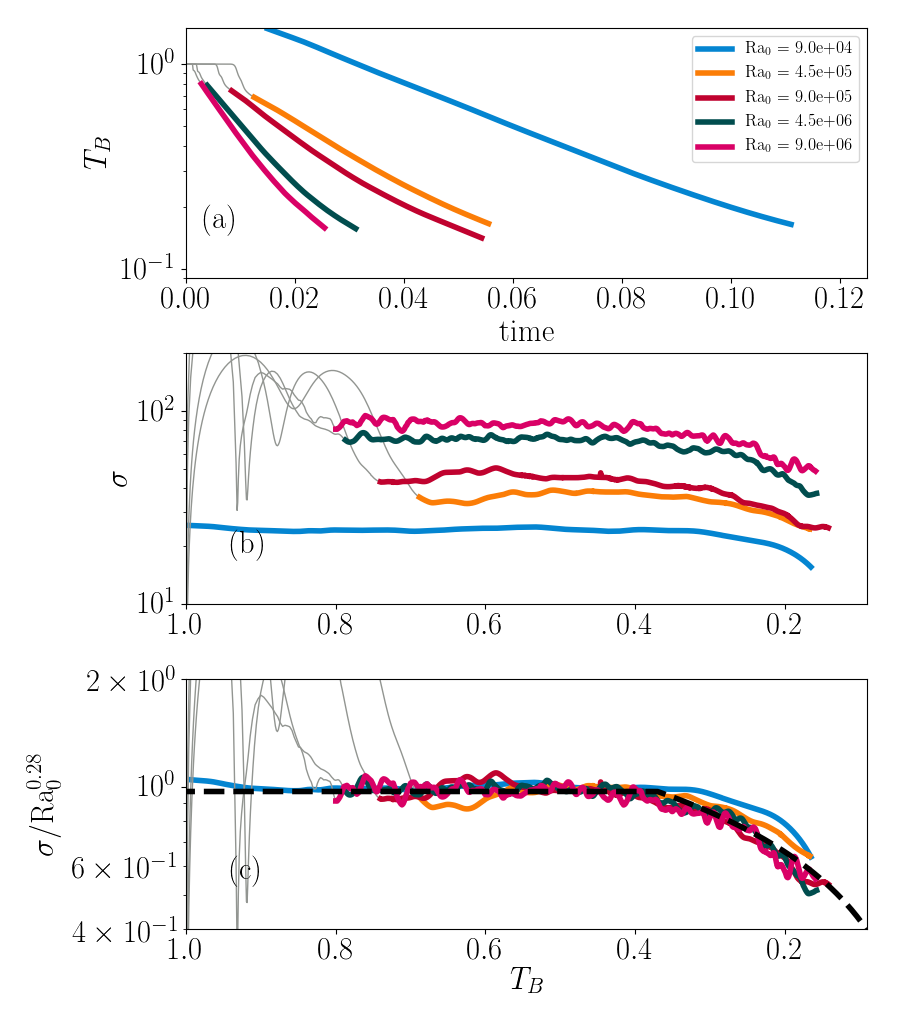}
\caption{Panel (a) is a plot of the bottom water temperature $T_B$ with time for all numerical runs with a nonlinear EOS. Panel (b) is a plot of the instantaneous decay rate ($\sigma$) as a function of $T_B$. Panel (c) is a plot of the scaled decay rates ($\sigma$), which collapse onto a single curve. We include the scaling \eqref{eqn::SigmaScl} as a dotted black line in panel (c).  We plot in grey the initial transition period before reaching a quasi-equilibrium state. Note that the x-axis has been reversed in panels (b),(c), with $\TB$ decreasing towards the right such that it reads in the same direction as time. }
\label{fig::DecayRate}
\end{figure}

  As discussed above, $\TB$ initially decays exponentially with time. Figure \ref{fig::DecayRate}(a) is a plot of $T_B$ as a function of time on a semi-log axis for all five of the nonlinear EOS simulations listed in Table \ref{Table::NumParams}. We compute the instantaneous decay rate as \[\sigma = \frac{-1}{\TB} \deriv{t}{\TB},\] and we observe (Figure \ref{fig::DecayRate}(b)) that, after an initial transient, $\sigma$ is nearly constant over a range of $T_B$, with $\sigma \gg 1$ ($\TB\gtrsim \TBTV$). 
  During this period, we perform a linear regression to determine the $\Ra$-number dependence of the flow. Figure \ref{fig::DecayRate}(c) is a plot of $\sigma$, scaled by $\Ra^{0.28}$, which collapses the data for all of the simulations with a nonlinear EOS. For lower values of $\TB$ ($\TB\lesssim \TBTV$), $\sigma$ decreases with $\TB$.  Once the system has achieved a quasi-steady state, we fit piecewise-curves to the different cases and approximate 
\begin{gather}
\sigma =  0.97 \Ra^{0.28} \begin{cases}
1, & \TB \ge \TBTV\\
 \left(\frac{\TB}{\TBTV}\right)^{0.63}, & \TB < \TBTV
\end{cases}
. \label{eqn::SigmaScl}
\end{gather}
We find a regime change at $\TB\approx\TBTV$, between a constant exponential decay ($\TB\gtrsim \TBTV$), and when $\sigma$ rapidly decreases ($\TB\lesssim\TBTV$). We note that the model presented in \S \ref{sec::ScalingLaws} is applicable only for $\sigma\gg 1.$ As $\TB\to 0$, $\sigma$ decreases and higher-order corrections are increasingly significant. It is also worth noting that Case 1: $\Ra=9.0\times 10^4$ has the lowest value of $\sigma$ and $\Ra$, and does not collapse as well as the other cases. 

 In connection with the linear EOS, we might expect that $\sigma$ should scale with an effective Rayleigh number ($\RaE$) (as was highlighted in \cite{Anders_2020}). This is partially correct. 
To justify this statement, we approximate equation \eqref{eqn::SigmaScl} as:
\begin{gather}
    \frac{\sigma}{\sigma_0} \approx \min \left\{ \RaEM^{0.28}, \RaE^{0.28}\right\} , \qquad \RaEM = \RaE(\TB = \TBT),
\end{gather}
where $\sigma_0 = 0.97 \TBT^{-0.63}$, and $\TBT = \TBTV$. 
That is, the temperature decay rate $\sigma$ does increase with $\RaE$ below some threshold value. For large enough $\RaE$, the convection is sufficiently turbulent that $\sigma$ remains constant. Thus, the convection is self-limiting. We believe this results from the increased diffusion at the top boundary due to the stable thermal layer that is not present in a linear equation of state. Importantly, as highlighted in \citet{Toppaladoddi2017,wang2019}, a third parameter must be defined to characterize the convection. We define that third parameter as $\TB.$

\begin{figure}
\centering
\includegraphics[width=\textwidth]{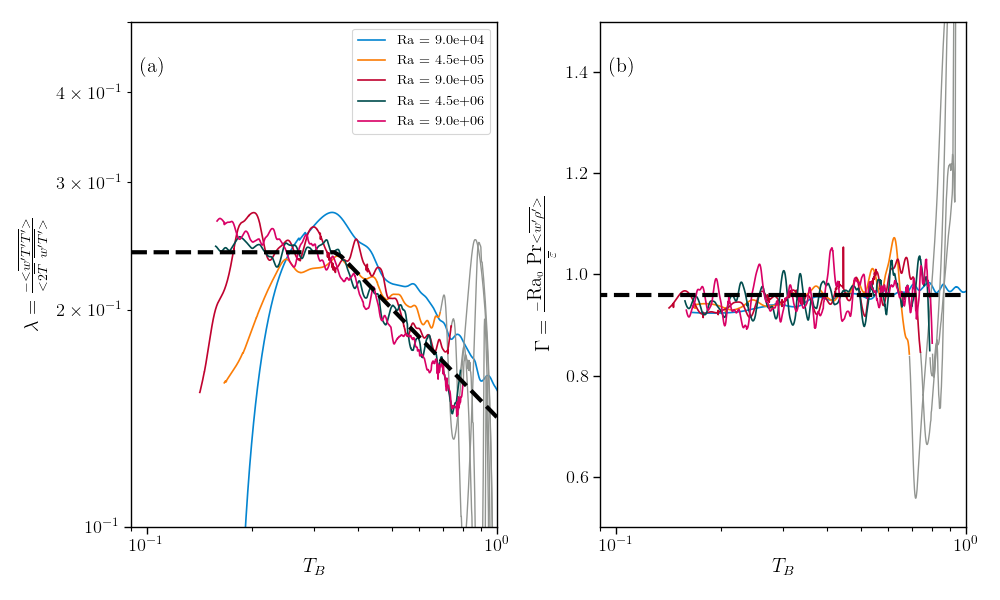}
\caption{Plot of (a) the buoyancy flux ratio $\lambda$ and (b) the mixing coefficient $\Gamma$ as a function of $\TB$. We include the fit of equation \eqref{eqn::LambdaFit} as a dashed line in panel (a). We further include a dashed line at $\Gamma = 0.96$ in panel (d). We plot in grey the initial transition period before reaching a quasi-equilibrium state.}
\label{fig::MixingCoefficientDivision}
\end{figure}

The mixing coefficient $\Gamma$ and buoyancy flux ratio $\lambda$, defined in equations \eqref{eqn::Lambda}, are also functions of the nondimensional parameters. Figure \ref{fig::MixingCoefficientDivision} is a plot of (a) $\lambda$ and (b) $\Gamma$ as a function of $\TB$. We find that $\lambda$ increases as $\TB$ decreases. Fitting $\lambda$ as a function of $\TB$ with a piecewise-functions, we estimate that while in quasi-steady state,  
\begin{gather}
\lambda =   0.24  \begin{cases}
\TB^{-\frac 12}, & \TB \ge 0.35\\
1, & \TB < 0.35
\end{cases}
. \label{eqn::LambdaFit}
\end{gather}
As the data is noisy, our estimated power-law dependencies are rough approximations, preferentially weighting the fit values of the higher $\Ra$ cases. We anticipate future work to illuminate a theoretical prediction for the appropriate scaling laws. 
In addition, it is important to note that Case 1: $\Ra=9.0\times 10^{4}$ does not follow the trend of the other cases. For Case 1, the low Rayleigh number results in a significant diffusive contribution to the total heat flux and is in a weakly unstable regime. As in equation \eqref{eqn::SigmaScl}, we find that there is a regime change that occurs at $\TB\approx\TBTV$.
Similarly, we find that  
\[\Gamma\approx 0.96\]
for all $\Ra$, though large fluctuations are present. This value of $\Gamma$ indicates that the rate of viscous dissipation is nearly equal to the vertical buoyancy flux. In steady-state, by definition, $\Gamma=1$ which is consistent with our observation that that the cooling box is nearly in steady state.

\begin{figure}
\centering
\includegraphics[width=0.6\textwidth]{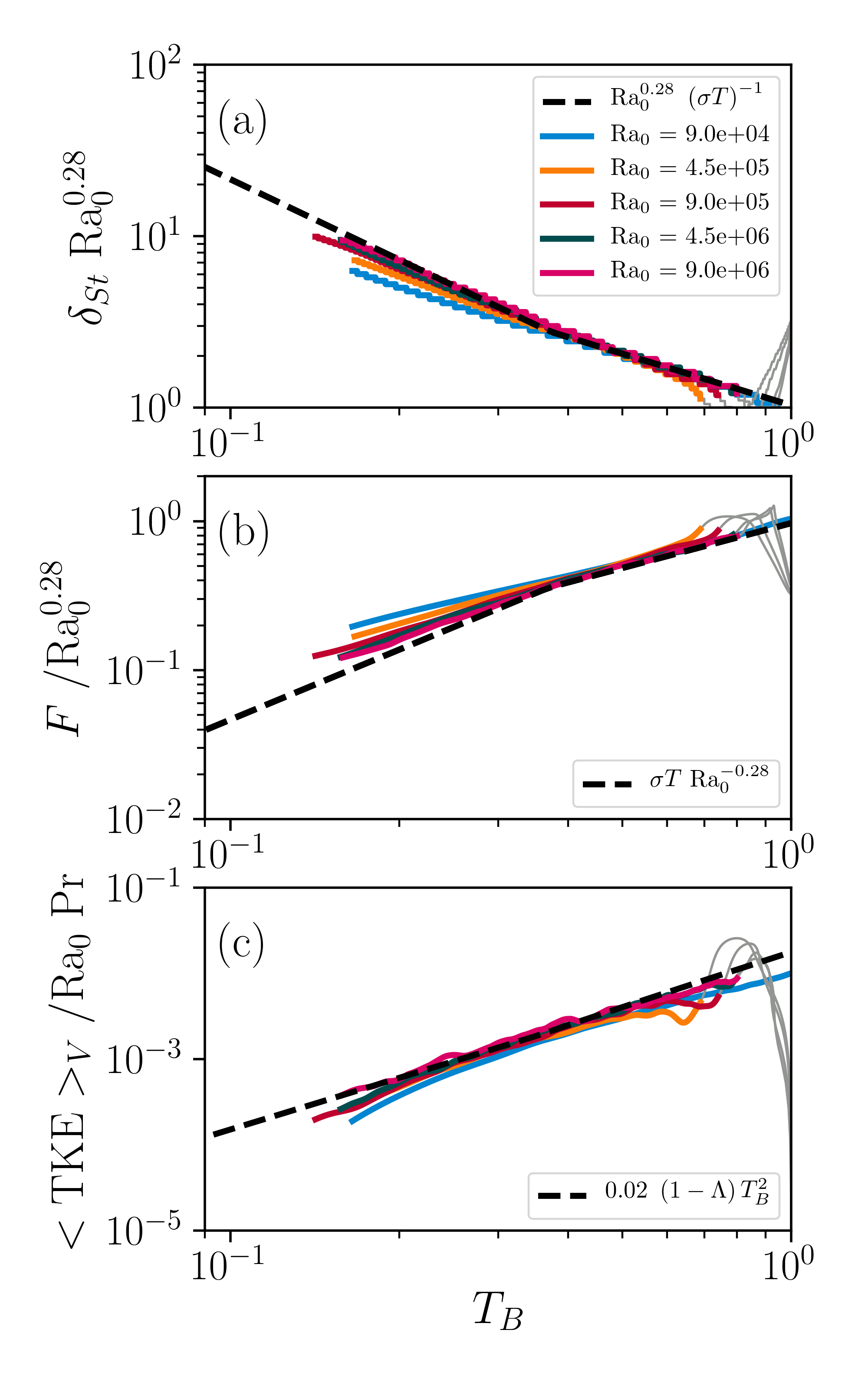}
\caption{A plot of the scaled (a) boundary layer thickness $\delta_{St}$, (b) surface heat flux $F$, and (c) turbulent kinetic energy density $\TKE$, as a function of $\TB$. The black dashed lines denote the predictions of \eqref{eqn::DelScaling} and \eqref{eqn::TKEScaling}.}
\label{fig::SelfSimilarCollapse}
\end{figure}

\newpage

\subsection{Model Agreement}
Can we predict the vertical heat transport and kinetic energy produced by the turbulent convection? In \S\ref{sec::ScalingLaws}, we derived scaling laws for $\delST$, $F$, and $\TKE$ as a function of $\Ra, \Pr,$ and $\TB$. 
Figure \ref{fig::SelfSimilarCollapse} is a comparison plot between the model equations \eqref{eqn::DelScaling} and \eqref{eqn::TKEScaling} and the numerically computed (a) $\delST$, (b) $F$, and (c) $\TKE$. The model agrees well with the data from the direct numerical simulations. We have scaled the y-axes by the appropriate powers of $\Ra$ and $\Pr$. Note that the scaling coefficient $0.02$ in panel (c) is equivalent to $\Gamma=0.96$. The parameters $\sigma$ and $\Lambda$ are defined using the fit equations \eqref{eqn::SigmaScl} and \eqref{eqn::LambdaFit}, respectively.

We further find that as the bottom water $\TB\to 0$, the simulated values of $\delta$ and $F$ appear to diverge from the model. We recall the model equations \eqref{eqn::DelScaling} are valid for $\sigma \gg 1$. As the decay rate $\sigma$ increases with $\Ra$, the larger values of $\Ra$ show better agreement between the model and the simulated data. Similarly, Case 4 - $\Ra = 9\times 10^4$ exhibits the largest deviation from the model fit as it also is the lowest $\Ra$ simulation. As the bottom water temperature decreases ($\TB\to 0$), $\sigma$ decreases rapidly. The first-order approximation (equation \eqref{eqn::DelScaling}) is not expected to perform well in that limit. Higher-order corrections can account for this discrepancy, but that is outside of the scope of this article.

The flow transition that occurs at $\TB\approx \TBTV$ (see equations \eqref{eqn::SigmaScl} and \eqref{eqn::LambdaFit}) results in a `kink' in the modelled predictions, as seen in Figure \ref{fig::SelfSimilarCollapse}. We do not yet have a prediction for this limiting temperature value of $T_{max} = \TBTV$. However, this transition is important in the evolution of $\delST$, $F$ and, to a lesser extent, $\TKE$.

\section{Conclusions}\label{sec::Conclusions}

We considered a box of warm fluid, cooled from the surface through a fixed-temperature boundary condition. In this box, the density was quadratic with temperature. The top boundary temperature and the initial domain temperature were selected to be on opposite sides of the temperature of maximum density, leading to the generation of convection and the formation of an upper stable layer and a lower convectively unstable layer. As the convection mixed the lower-layer fluid, its near homogeneous temperature decreased ($\TB\to0$).

We developed a model for our system and scaling laws for $\delST$, $F$, and $\TKE$, 
\begin{gather}
\delta_{St} \sim \frac{1}{\sigma T_B}, \qquad  F = \frac{1}{\delta_{St}} \sim \sigma T_B, \qquad \sigma \gg 1 \label{eqn::Model1} \\
\Vavg{\TKE} \sim \frac 12 \left( \Gamma^{-1} - 1 \right) \left( 1 - \Lambda \right) \Ra \Pr \TB^2. \label{eqn::Model2}
\end{gather}
Analyzing the numerical simulations, we determined the $\Ra$ and $\TB$ dependence of the decay $\sigma$ and the integrated buoyancy flux ratio $\Lambda$. We showed that the mixing coefficient $\Gamma$ was effectively constant over the whole range of parameters considered here. Once determined, the model equations agreed well with the numerically computed $\delST$, $F$, and $\TKE$. 

\vspace{6pt}

\setlength{\fboxrule}{1.5pt}
\noindent \fbox{\begin{minipage}{\textwidth}

\colorlet{shadecolor}{gray!20}
\begin{shaded*}
\vspace{-4pt}
This paper's main goal was to understand how convection is changed when the equation of state is nonlinear. We have shown that :
\begin{enumerate}
    \vspace{6pt}
    \item \ The surface heat flux is dramatically different, in both magnitude and parameter dependence, between a nonlinear and linear equation of state.
    \vspace{4pt}
    \item \ In addition to the Rayleigh and Prandtl number, convection with a quadratic equation of state depends on a third nondimensional parameter, the bottom water temperature $\TB$.
    \vspace{4pt}
    \item \ Our model \eqref{eqn::Model1}-\eqref{eqn::Model2} accurately predicts the heat flux ($F$), boundary layer thickness ($\delST$), and turbulent kinetic energy ($\TKE$) based on these three parameters.
    \vspace{-4pt}
\end{enumerate}
\end{shaded*}

\end{minipage}}

\vspace{6pt}

This work is a crucial step towards understanding how a nonlinear equation of state modifies convection. The quadratic equation of state limits the vertical heat flux and the kinetic energy, compared to a linear equation of state, and depends on an additional nondimensional parameter $\TB$. We are in the process of constructing a Cold Convection Facility, capable of convectively cooling the surface of a fresh body of water to run complementary laboratory experiments. This work provides a framework, including the essential parameters and model considerations, to understand that much more complicated system and, subsequently, freshwater systems in the environment. 


\textbf{Declaration of Interests:} The authors report no conflict of interest.

\section*{Acknowledgements}
We want to thank Andrew Wells, Hugo Ulloa, and Louis-Alexandre Couston for their feedback on this work. This work was funded in part by the Natural Sciences and Engineering Research Council of Canada, the Isaak Killam Trust, and Syncrude Canada Ltd.


\appendix

\section{Derivation of the scaling law for $\delST$}\label{App::DerivDelta}

First, assuming a piecewise-linear profile for $\mT$, we can establish that 
\begin{gather}
 \delBL = \left(1+\TB\right) \delST .\label{eqn::delBL}
\end{gather}
The rate of change of the total temperature within the domain is then, 
\begin{align}
\deriv{t}{} \int_0^1 \bar T \dz 
&=\deriv{t}{} \left( \int_0^{1 - \delta_{BL}} T_B \dz + \int_{1-\delta_{BL}}^1\left(  -1 - \frac{1 }{\delta_{St}} \left(z - 1 \right) \right) \dz\right)  \\
  &= \deriv{t}{T_B} \left(1-\delta_{BL}\right) -  \frac 12 \frac{\delta_{BL}^2}{\delta_{St}^2} \pderiv{t}{\delta_{St}} 
\end{align}
Noting that the top temperature gradient prescribes the rate of change of heat within the domain, and including \eqref{eqn::delBL}, we arrive at 
\begin{gather}
\deriv{t}{T_B} \left(1-\left(1 + T_B\right)\delta_{St}\right) -  \frac{\left(1 + T_B\right)^2}{2} \pderiv{t}{\delta_{St}} = -\frac{1}{\delta_{St}}
\end{gather}

If we further evaluate 
\[  \frac{d \TB}{dt} \sim -\sigma \TB, \qquad \delST\ll1, \qquad \sigma\gg1,\]
then to leading order 
\begin{gather}
\delta_{St} \sim \frac{1}{\sigma T_B}, \qquad \sigma\gg1.
\end{gather}

\section{Derivation of the scaling law for $\TKE$}\label{App::DerivTKE}

We first recall the Reynolds decomposition 
\[
\mathbf{u} = \mathbf{0} + \mathbf{u}', \qquad T = \mT + T'. 
\]
The density flux is then written out as 
\[ 
\rho = - T^2 = - \left( \mT^2 + 2 \mT T' + T^\prime T^\prime\right) \implies \overline{w^\prime \rho^\prime} =  - 2 \Havg{w^\prime T^\prime} \ \Havg{T} - \Havg{w^\prime T^\prime T^\prime}.
\] 
For $z<1-\delBL$, the temperature stratification is well mixed, and thus 
\[ \Havg{T'w'} = -\deriv{t}{T_B}z = \sigma T_B z, \quad z<1-\delBL \implies \Vavg{2 \Havg{w^\prime T^\prime} \ \Havg{T}} \approx \int_0^{1} 2 \sigma T_B^2 z \dz = \sigma T_B^2, \quad \delBL\ll1.\]
The approximations can be formalized by performing the full integrals and expanding the $\delBL$ as a perturbation series in $\frac{1}{\sigma}$. Therefore, the TKE equation reduces to 
\begin{gather}
    \deriv{t}{\Vavg{\TKE}} = -\left(\Gamma^{-1} - 1\right) \left( 1 - \lambda\right) \sigma \TB^2.
\end{gather}

The final step in deriving equation \eqref{eqn::TKEScaling} is to use self-similarity to determine, 
\begin{gather}
\deriv{t}{\Vavg{ \TKE (\Ra,\Pr,\TB)}} = \deriv{\TB}{\Vavg \TKE} \deriv{t}{\TB}.
\end{gather}
Substituting in for $\Gamma$ and $\Lambda$ and integrating with respect to $\TB$, results in 
\begin{gather}
\Vavg{\TKE} \sim \frac 12 \left( \Gamma^{-1} - 1 \right) \left( 1 - \Lambda \right) \Ra \Pr T_B^2, \qquad \sigma \gg 1,
\end{gather}
where $\Gamma$ is assumed constant and
\begin{gather}
    \Lambda = \frac{2}{\TB^2} \int_{\TB} T' \lambda(T') \ dT'.
\end{gather}

\section{Linear Stability Analysis} \label{App::LinStab}

The initial evolution of the temperature profiles described in \S \ref{sec::NumSim} is diffusive. For a deep box, the diffusive temperature solution is 
\begin{gather}
\overline T = -1 - \left(1+\TB\right) \hbox{erf} \left( \frac{z}{\delta}\right), \qquad
\delta = \sqrt{4 t},\label{eqn::delta}
\end{gather}
where $\TB$ is fixed for the purposes of this stability analysis. 
Here, as we have a deep box, we will redefine $z=0$ as the top boundary with the domain of interest below. In this analysis, we need to include diffusion of this background profile in the linear stability. We follow the approach of \cite{Nijjer2018} and define the similarity variable 
\begin{gather}
\xi = \frac{z}{\sqrt{t}},
\end{gather}
such that the background density profile is
 \begin{gather}
\bar T =-1 - \left(1+\TB\right) \hbox{erf} \left( \frac{\xi}{2}\right).
\end{gather}
We want to know the growth rate of infinitesimal modal perturbations to the diffusive background state. The linear vertical velocity ($w_\epsilon$) and temperature ($T_\epsilon$) perturbations are assumed to have the form:
\begin{gather}
\begin{bmatrix}
 w_\epsilon \\   T_\epsilon
\end{bmatrix} = \begin{bmatrix}
\hat w \\  \hat T
\end{bmatrix}\tau(t) \exp \left[ i \mathbf{k}\cdot \mathbf{x}\right].
\end{gather}
 Following the approach of \cite{drazin}, we linearize the equations of motion \eqref{eqn::momentum}-\eqref{eqn::DivFree}, which will result in a linear eigenvalue problem of the form:
 \begin{gather}
\mathbf{\underline{A}} \begin{bmatrix}
\hat w \\  \hat T
\end{bmatrix} = \lambda \mathbf{\underline{B}} \begin{bmatrix}
\hat w \\  \hat T
\end{bmatrix},
\end{gather}
 where $\mathbf{\underline{A}},\mathbf{\underline{B}}$ are matrix operators and $\lambda  = \frac{1}{\tau} \frac{d \tau}{dt} \bigg|_{t=t0}$ is the growth rate of the perturbations at some time $t0$. 
The solutions to these eigenvalue equations are a function of five parameters: $t0,k$, Ra, Pr, and $T_B$. 
We solve these equations using an in-house built eigenvalue solver using Chebyshev differentiation matrices. We found that 50 grid points were sufficient to determine the growth rate of the system. We impose the same top and bottom boundary conditions as in the numerical simulations. Spurious solutions that did not satisfy the appropriate boundary conditions were removed.

\begin{figure}
\centering
\includegraphics[width=\textwidth]{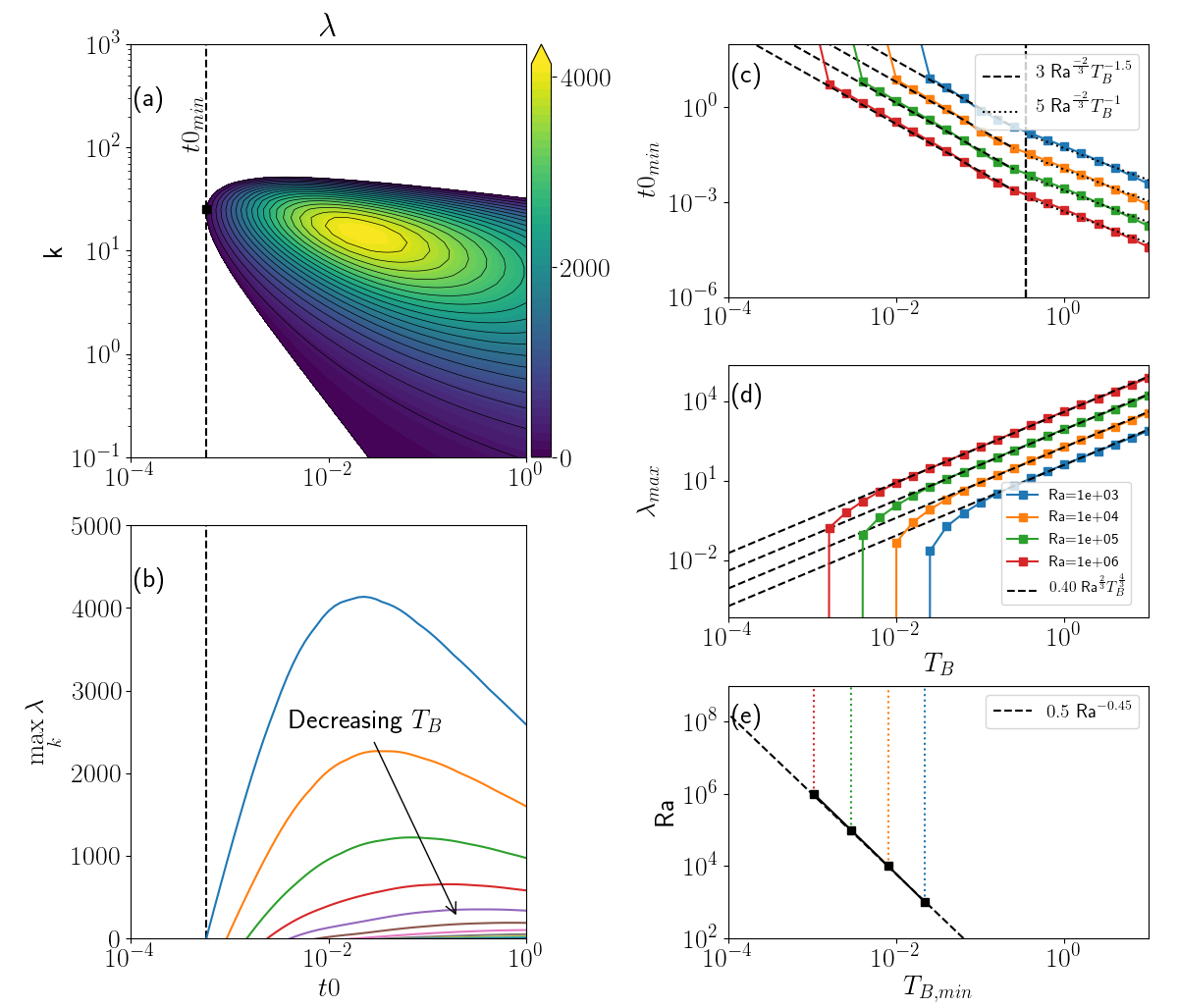}
\caption{Panel (a) is a contour plot of the growth rate ($\lambda$) of the linear instability as a function of $t0$ and wavenumber ($k$). Panel (b) is a plot of the maximum growth rate for all $k$ as a function of $t0$ for decreasing $\TB$. Both panel (a) and (b) are computed for $\hbox{Ra}=10^6$ and $\Pr=9$. Panel (c) is a plot of the minimum time to instability $t0_{min}$ below which the system is stable as a function of $\TB$. We include a vertical line at $\TB=0.35$ to show a similar regime change to the nonlinear dynamics. Panel (d) is a plot of the maximum growth rate for all $k$ and $\delta$ as a function of $\TB$. Panel (e) is a plot of the minimum $\TB$ below which the system is linearly stable versus $\hbox{Ra}$.}
\label{fig::LinStabGrowth}
\end{figure}

Figure \ref{fig::LinStabGrowth}(a) is a plot of the growth rate ($\lambda$) of the linear instability as a function of wavenumber $k$ and $t0$ for $\hbox{Ra}=10^6, \Pr=9,$ and $\TB = 1$. We first notice that there exists a minimum $t0_{min}$, below which the system is linearly stable. For $t>t0_{min}$, the system is unstable for a finite range of wavenumbers and a peak $\lambda$ at $t\gg t0_{min}$. Figure \ref{fig::LinStabGrowth}(b) is a plot of the maximum $\lambda$ over all wavenumbers, for each time $t0$ at different $\TB$. The maximum $\lambda$ decreases with $\TB$ until the system becomes stable at finite $\TB$.

By computing $t0_{min}$, we determine the earliest time at which the diffusive system becomes linearly unstable. Figure \ref{fig::LinStabGrowth}(c) is a plot of $t0_{min}$ as a function of $\TB$ for different Ra. Fitting the data, we find that 
\begin{gather}
t_0 = \hbox{Ra}^\frac{-2}{3}\begin{cases} 
3 \TB^{-1}, & \TB<0.35\\
5 \TB^{\frac{-3}{2}}, & \TB\ge0.35
\end{cases}\label{eqn::LinStabDelMin}
\end{gather}
This fit is included as dashed/dotted lines in Figure \ref{fig::LinStabGrowth}(c). As with the nonlinear simulations, we observe that there exists a critical regime change that occurs around $T_B\approx 0.35$. Further, there is a minimum $\TB$ [$T_{B,min}$], where the system is linearly stable for all time and wavenumbers. Figure \ref{fig::LinStabGrowth}(d) is a plot of the maximum growth rate ($\lambda_{max}$) for all $k$ and $t0$ as a function of $\TB$. For $\TB$ close to 1, the growth rate of the system follows 
\begin{gather}
\lambda_{max} \approx 0.4 \hbox{Ra}^\frac 23 \TB ^\frac 43 = 0.4 \RaE^\frac 13.
\label{eqn::LinSigScaling}
\end{gather}
Again, this fit is included as dashed lines in Figure \ref{fig::LinStabGrowth}(d).
As $\TB\to T_{B,min}$, $\lambda_{max}$ diverges from the fit \eqref{eqn::LinSigScaling}, decreasing rapidly to 0.

For a linear EOS, we know that there is a minimum Rayleigh number below which the system is stable. Similarly, we have found a minimum condition for instability with a quadratic EOS that depends on both $\hbox{Ra}$ and $\TB$. Figure \ref{fig::LinStabGrowth}(e) is a plot of $T_{B,min}$ for different values of $\hbox{Ra}$. Note that this panel has been rotated so that $T_{B,min}$ can be easily compared with panels (c) and (d). We estimate
\begin{gather}
T_{B,min}\approx 0.5 \hbox{Ra}^{-0.45}. \label{eqn::LinMinTB}
\end{gather}
Notice that in terms of $\RaE$, this suggests that the convection is stable where $\RaE\lesssim \frac{1}{4}$. There exists a minimum stability condition below which the system is linearly stable for these diffusive temperature profiles for all time.

For $\RaE$ larger than this minimum stability criterion, the analysis highlights that there still exists a $t0_{min}$ (or equivalently a minimum interface thickness), below which the system remains linearly stable. For $t>t0_{min}$, perturbations about the base state will grow and result in convective mixing of the temperature field. We can compare this minimum time with the numerical simulations.

\begin{figure}
\centering
\includegraphics[width=\textwidth]{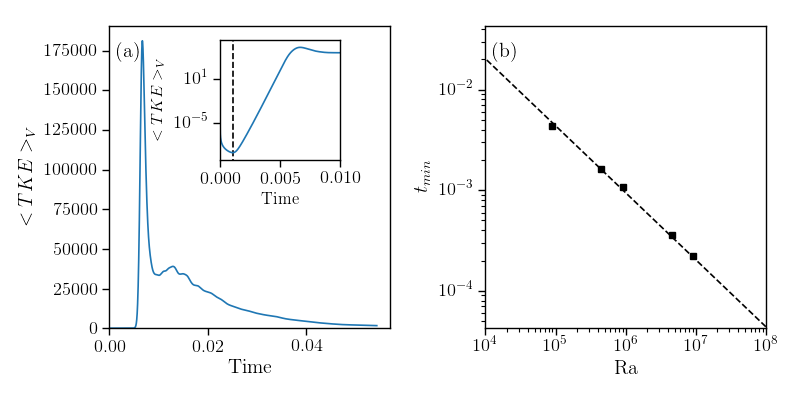}
\caption{(a) A plot of the kinetic energy as a function of time for Case 3: $\Ra=9.0\times10^5$. The inset presents a subset of the data on a log-linear axis to show the initial TKE growth. The vertical dashed line indicates the time of minimum kinetic energy ($t_{min}$). Panel (b) is a plot of $t_{min}$ as a function of Rayleigh number (Cases 1,2,3,4, and 5). The scaling law \eqref{DNS::ScalingLaw} is included as a dashed black line in panel (b). }
\label{fig::t0_vs_ra}
\end{figure}

Figure \ref{fig::t0_vs_ra}(a) is a plot of the volume integrated turbulent kinetic energy density ($\TKE$). As with the linear stability, the initial temperature stratification is stable, such that $\TKE$ initially decays (see Figure \ref{fig::t0_vs_ra}(a) inset). Once the top boundary layer is sufficiently deep, flow instability results in a large peak in the kinetic energy, which quickly decays back to an equilibrium value. From there, the $\TKE$ slowly decays. 
We define the time when the kinetic energy reaches a minimum (vertical dashed line in Figure \ref{fig::t0_vs_ra}(a) inset) as $t_{min}$. Note that this is only an approximation for the time when the system becomes linearly stable. Figure \ref{fig::t0_vs_ra}(b) is a plot of $t_{min}$ as a function of the Rayleigh number. A fit of the data to the expected power-law suggests that, for the numerical simulations,
\begin{gather}
t_{min} \approx 4.4 \Ra^\frac{-2}3, \label{DNS::ScalingLaw}
\end{gather}
which is close to the expected from linear theory from \eqref{eqn::LinStabDelMin} with $\TB=1$.


	\section{Resolution and Domain Size Verification} \label{App::Res}
	
	\subsection{Domain Dependency}
\begin{figure}
	\centering
	\includegraphics[width=10cm]{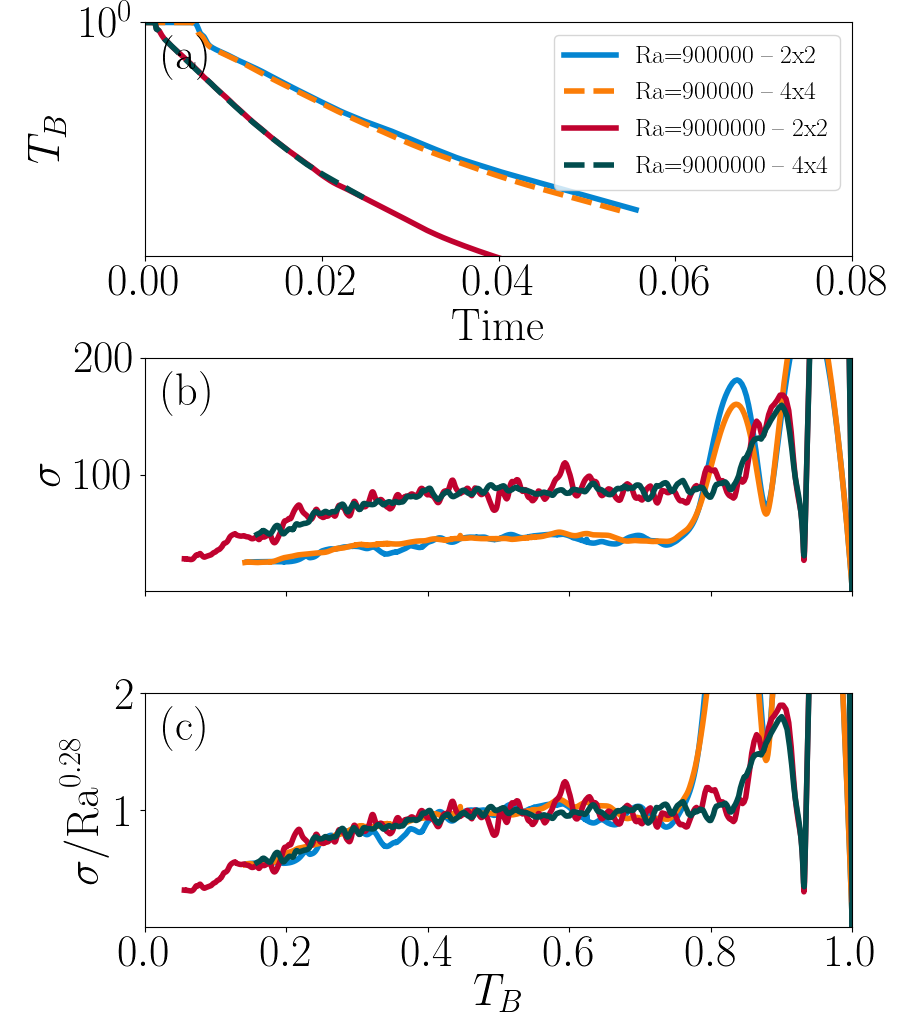}
	\caption{Similar to Figure \ref{fig::DecayRate}, we demonstrate that the decay rate of change is similar for different sized domains.     }
	\label{fig::DecayRate_DomDep}
	\end{figure}

    As mentioned in the text, we verified that the present results are not dependent on the box size of the numerical simulations. Figure \ref{fig::DecayRate_DomDep} is similar to Figure \ref{fig::DecayRate}, with data from two different sized domains. We show that when we double the numerical domain, the temperature decay rate is nearly identical, despite the different initial random noise. 
	
	\subsection{Resolution Sufficiency}
\begin{figure}
	\centering
	\includegraphics[width=10cm]{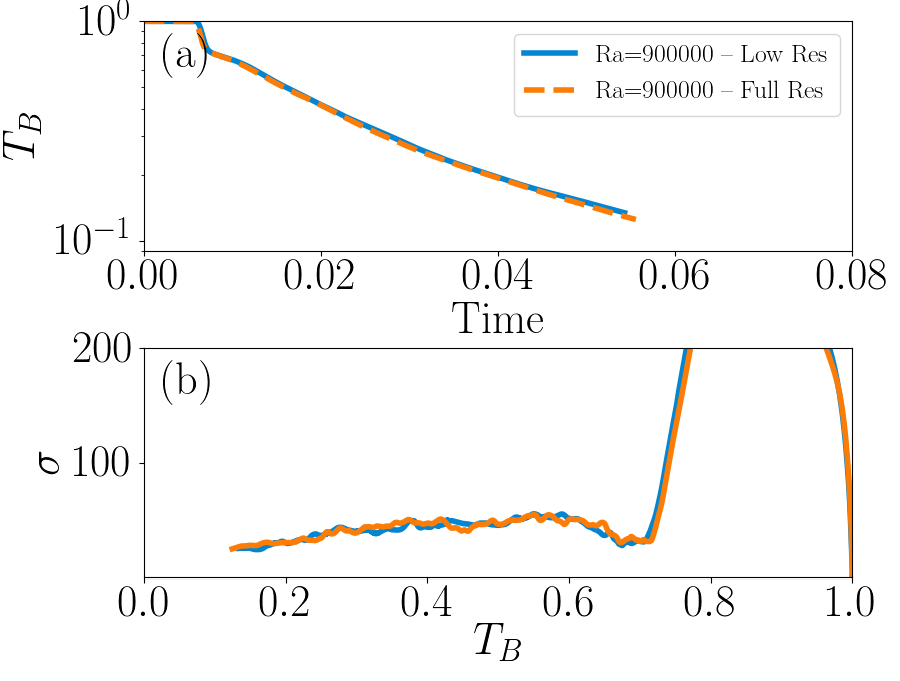}
	\caption{Similar to Figure \ref{fig::DecayRate}, we demonstrate that the simulation resolution is sufficient to resolve the rate of change of temperature.    }
	\label{fig::DecayRate_ResDep}
	\end{figure}

	To demonstrate that the numerical simulations are indeed resolved, we have run an additional low-resolution simulation. We re-run Case 3 $-- \Ra = 9\times 10^5$, but with a grid resolution of Nx $\times$ Ny $\times$ Nz = $128 \times 128 \times 128$, or half the resolution in each direction. Figure \ref{fig::DecayRate_ResDep} is a comparison plot between the full-resolution Case 3, and the low-resolution run described here. We observe that the rate of change of temperature is essentially unchanged with one-eighth the number of grid points. 
	
\bibliography{CoolingBox}
\bibliographystyle{apalike}

\end{document}